\documentclass[11pt,a4paper]{article}
\usepackage[utf8]{inputenc}
\usepackage[letterpaper,left=16mm,right=16mm,top=20mm,bottom=35mm]{geometry}

\usepackage{libertine}
\usepackage[T1]{fontenc}
\usepackage{microtype}

\usepackage{enumitem}
\setlist[itemize]{leftmargin=20pt}
\usepackage[american]{babel}
\usepackage{multicol}

\usepackage[dvipsnames]{xcolor}
\usepackage{url}
\usepackage[
  colorlinks=true,
  allcolors=NavyBlue,
]{hyperref}

\usepackage{titlesec}
\titlespacing{\paragraph}{%
  0pt}{%
  0.5\baselineskip}{%
  1em}

\titleformat*{\paragraph}{\normalsize\em}

\makeatletter
\renewcommand{\maketitle}{\bgroup\setlength{\parindent}{0pt}
  \begin{flushleft}
    {\bf\LARGE\@title}\\[16pt]
    \@author
  \end{flushleft}\egroup
}
\makeatother

\expandafter\def\expandafter\UrlBreaks\expandafter{\UrlBreaks\do\/\do\*\do\-\do\~\do\'\do\"\do\-}
\usepackage[most]{tcolorbox}
\usepackage{xspace}
\newcommand{\ourbench}{BrokenBench\xspace}

\tcbuselibrary{breakable}
\newtcolorbox{runexample}[1][]{%
  colback=black!3,
  colframe=black!50,
  coltitle=black,
  colbacktitle=black!12,
  fonttitle=\bfseries\sffamily\small,
  title={Example: #1},
  boxrule=0.6pt,
  arc=0pt,
  left=4pt, right=4pt, top=4pt, bottom=4pt,
  before skip=16pt, after skip=16pt,
  fontupper=\small\sffamily,
  breakable, enhanced
}

\title{Measuring Security Without Fooling Ourselves:\\ Why Benchmarking Agents Is Hard}

\author{Sahar Abdelnabi$^1$, Chris Hicks$^2$, Konrad Rieck$^3$, and Ahmad-Reza Sadeghi$^4$\\[10pt]
  \normalsize $^1$~ELLIS Institute Tübingen \& MPI-IS \& T\"ubingen AI Center, Germany\\
  \normalsize $^2$~The Alan Turing Institute, London, UK\\
  \normalsize $^3$~BIFOLD \& Technische Universität Berlin, Germany\\
  \normalsize $^4$~Technische Universität Darmstadt, Germany}

\date{\relax}

\begin{document}

\thispagestyle{empty}

\maketitle

\section*{Abstract}

The benchmarks used to evaluate AI agents in security-critical roles suffer from crucial weaknesses. Building on recent empirical evidence, we characterize three core challenges that undermine security evaluations: \emph{benchmark vulnerabilities}, \emph{temporal staleness}, and \emph{runtime uncertainty}. We then outline practical directions toward building more robust and trustworthy evaluation frameworks.

\bigskip
\begin{multicols}{2}

	\section{Introduction}
	Evaluating AI agents in security settings presents a fundamental challenge. The very capability we seek to measure, adversarial reasoning, can undermine the evaluation itself. An agent tasked with discovering vulnerabilities may also exploit weaknesses in the benchmark environment instead of solving the intended task, creating a mismatch between measured performance and model capability.

	Addressing this challenge requires a shift in perspective: benchmarking AI agents is not only about measuring efficiency or task performance, but is also itself a systems security problem. Unlike traditional benchmarking, where it might be adequate to assume a trusted and static evaluation setup, researchers running experiments with security agents must treat the benchmark as an adversarially exposed system that can be attacked and manipulated.

	Recent work has produced a growing ecosystem of security benchmarks. Cybench evaluates agents on professional-level capture-the-flag challenges~\cite{cybench}, CyberGym tests vulnerability reproduction across real-world CVEs~\cite{cybergym}, and PentestGPT established the first systematic evaluation of LLM-driven penetration testing~\cite{pentestgpt}.
	Despite this progress, none of these benchmarks are designed to withstand the adversarial capabilities they aim to measure. Consequently, empirical evidence already shows that agents can exploit benchmark weaknesses to achieve near-perfect scores without solving the intended tasks~\cite{berkeley}.

	To make this challenge concrete, we introduce a simplified running example. An agent is evaluated on the fictional vulnerability discovery benchmark \emph{\ourbench}, where it is given access to programs running inside isolated containers. The agent can inspect source code, execute binaries, and use analysis tools to identify vulnerabilities. The evaluation environment of \ourbench includes not only these programs, but also the surrounding infrastructure, such as the container runtime, configuration files, orchestration logic, and scoring mechanisms for grading.

	Crucially, this environment is part of the attack surface for the agent. Rather than analyzing the intended target programs, the agent may thus exploit the infrastructure of \ourbench, for example, by accessing hidden solution data, manipulating the scoring process, or escaping isolation boundaries. In such cases, the agent achieves high scores without solving the intended task. This illustrates the following core tension: in adversarial settings, it can be easier to exploit the benchmark than to solve it.

	Against this backdrop, this article highlights three critical limitations of current security benchmarks. First, \emph{benchmark vulnerabilities} allow agents to manipulate or escape evaluation environments, threatening result integrity. Second, \emph{temporal staleness} means that static benchmarks quickly become outdated in an evolving threat landscape. Third, \emph{runtime uncertainty}, arising from stochastic behavior, code generation, and external dependencies, introduces risks that are rarely captured in existing benchmarks.

	These challenges reveal that security benchmarking for AI agents is not only an evaluation problem but a systems security problem in itself. Robust assessment requires rethinking benchmarks as dynamic, adversarially robust, and continuously validated environments that reflect the realities of modern cybersecurity.

	\begin{runexample}[\ourbench]
		\emph{\ourbench} is a \emph{fictional} benchmark that presents an agent with different C programs running inside Docker containers, each containing a known vulnerability. The agent has shell access to each container, can read source code, compile and execute programs, and run analysis tools. For each program, it must identify the vulnerability type, locate the affected function, and produce a proof-of-concept exploit. Ground-truth labels, including the vulnerability type, location, and an exploit, are stored on the host outside the container.
	\end{runexample}

	\section{Benchmark Vulnerabilities}

	If an agent is designed to solve a task, such as finding complex vulnerabilities in software, it can also find shortcuts in the benchmark environment itself, creating a central paradox. That is, it may be \emph{easier} to break the environment than to solve the intended task.

	For example, in the BrowseComp benchmark which tasks agents with locating hard-to-find information online, agents that had processed approximately 30~million tokens began to recognize they were being tested and searched online for the benchmark specification itself~\cite{browsecomp}. A systematic audit found that agents could inject code via configuration files to rewrite test outcomes, inspect Git logs to reverse-engineer benchmark construction, and achieve 100\% exploitation rates on benchmarks including SWE-bench and Terminal-Bench~\cite{berkeley}.

	\begin{runexample}[Escaping the sandbox]
		The agent is tasked with finding a vulnerability.
		Instead of analyzing the source code, it probes the \ourbench environment and discovers that the container allows access to the hosts container management interface. Using this, the agent spawns a privileged container, reads the ground-truth data from the host filesystem, and returns the correct vulnerability for all challenges without ever analyzing them.
	\end{runexample}

	\columnbreak
	This unexpected outcome points to a requirement that security benchmarking has yet to recognize: the evaluation environment must be more secure than the target being tested.
	For example, if the target is a well-vetted cryptographic library and the benchmark runs on an average Linux system, the agent may find it easier to slip past the container than to dig out a subtle flaw in the library.

	\subsection{Inner and Outer Protections}

	In the context of this paradox we find it useful to distinguish between \emph{inner protections} (alignment-based constraints on the agent's behavior) and \emph{outer protections} (architectural constraints on the evaluation environment). Current benchmarks rely almost entirely on inner protections; trusting that the agent will attempt to solve the intended task rather than exploit its surroundings. This is the wrong assumption for security benchmarks where adversarial thinking is the capability being measured.

	Robust evaluation requires outer protections: hardware-enforced isolation, separate privilege domains for the agent and the answers, and verification of task completion. However, even well-designed sandboxes are not invulnerable. Container escapes are a known vulnerability class, and the sophistication of agent-generated exploits is increasing. The implication is that outer protections must be continuously monitored, hardened and tested against the agents they are meant to contain.

	\subsection{Canaries}

	Since the benchmark itself cannot be fully protected, we borrow a familiar idea from systems security: \emph{canary tokens}, hidden and randomized values embedded in the environment that an honest solver has no reason to observe. If an agent reproduces one of these canaries during evaluation, it has reached parts of the infrastructure that were never meant to be accessible and its score must not be trusted.
	This bridges to the classical concept of honeypots: detecting when something is not running as planned.
	Canaries are already used for detecting whether a model was trained on a particular dataset: unique canaries are injected into the data beforehand and the model is later checked for whether it reproduces them~\cite{satmlmi}. We believe this should be adopted as common practice, not just for training data, but for all sources that the agent may read at inference time that may enable benchmark cheating.

	\subsection{Deliberate Cheating}

	A recent audit~\cite{berkeley} revealed that benchmark exploitation is not limited to accidental shortcutting. Agents can be designed to \emph{deliberately} cheat. When given instructions encouraging exploitation of the evaluation environment, agents systematically identified and leveraged vulnerabilities in test harnesses, configuration files, and scoring mechanisms. This raises a practical question: can we use agents tasked with deliberate cheating to reveal limitations and shortcuts in the environment that can be exploited?

	In particular, computational cost can be treated as a coarse proxy for security.
	By setting a budget, e.g., ``if after \$10{,}000 the agent has not compromised the environment we consider it effectively secure
	'', technical hardening can be complemented with economic incentives. This is not a guarantee, but it raises the bar: if tasks are completed for less than the budget, and agents cannot cheat within the budget, then benchmark results are more likely to correspond to capability.

	\section{Temporal Staleness}

	What was secure yesterday is not secure today and threat models change over time. Vulnerabilities are discovered and patched continuously, new attack surfaces emerge, and defensive measures evolve. Yet a growing ecosystem of security benchmarks and evaluation frameworks use fixed datasets~\cite{cybench,cybergym,pentestgpt,agentauditor}.

	The consequences of stationary benchmarks are well understood in security. Intrusion detection datasets from the early 1990s remained in widespread use for over a decade despite becoming obsolete within two years of release~\cite{Tavallaee09}. We risk repeating this pattern: CyberGym's 1{,}507 vulnerabilities and Cybench's 40 CTF challenges
	will all age as the underlying software and agents are patched, new vulnerability classes emerge, and models are trained on the benchmark data.

	Another problem resulting from staleness is \emph{benchmaxxing}: fine-tuning models on benchmark formats and evaluation environments~\cite{hardtbench}. We also hypothesize that models trained on meta-knowledge about how benchmarks work, such as the format of multiple-choice security questions or the structure of CTF challenges, may show improved performance even without exposure to specific benchmark instances, regardless of their true capabilities. This could challenge the assumption that avoiding direct training on benchmarks is sufficient. The risk of benchmaxxing grows with the age of a benchmark and the volume of public material surrounding it.

	\begin{runexample}[Stale ground truth]
		\ourbench was released in 2024 with programs drawn from open-source projects. By 2026, most of the vulnerabilities have been patched upstream, and detailed write-ups appear in public CVE databases and blog posts. An agent trained in 2026 and evaluated on \ourbench would score considerably higher. The improvement reflects memorization of published fixes in the training data. Meanwhile, new vulnerability classes prevalent by 2026 are entirely absent from the benchmark.\\
	\end{runexample}

	\subsection{Dynamic Benchmarks}

	We argue that security benchmarks should function like a consumer price index: a representative basket of tasks that updates as the underlying security landscape changes. Just as a price index reflects how the actual mix of goods people buy evolves over time, a security benchmark should reflect how enterprise hosts, critical servers, and agents themselves are patched and upgraded over time.

	This requires a shift from static datasets to \emph{dynamic} benchmark environments. One approach draws on the model of fuzzing competitions, which run new rounds every few months with fresh challenges, allowing rankings to change as both the targets and the competitors evolve~\cite{fuzzbench21}. Another is to build benchmark systems that generate novel tasks from vulnerability databases (such as the CVE catalog) at regular intervals, producing fresh tasks of calibrated difficulty without human authorship of each instance.

	\subsection{Live Evaluation}

	A more radical approach is \emph{live evaluation}: benchmarking agents against real, currently-deployed systems rather than frozen snapshots. In fact, bug bounty platforms already operate this way, testing against production environments where the vulnerability landscape shifts daily. Adapting this model for agent evaluation would mean continuously refreshing target environments with current patch levels, real configurations, and newly disclosed vulnerabilities.

	The challenge is safety and reproducibility. Live evaluation sacrifices the controlled conditions that make benchmarks both relatively safe (i.e., to avoid doing harm to production systems) and comparable across time and agents. A hybrid approach may be more practical: maintaining a set of stable security tasks for longitudinal comparison while supplementing with a live set of tasks drawn from recent CVE disclosures, current CTF competitions, and LLM vulnerabilities. This mirrors how the NIST National Vulnerability Database provides a continuously updated stream that could, in principle, feed live benchmarks.

	\subsection{Generative Benchmarks}

	Agents could also be used to generate a live benchmarking environment that, rather than relying on human experts to curate benchmarks (a process that is expensive, slow, and does not scale); continuously tasks agents with discovering and patching vulnerabilities in the latest codebases, and then updating the benchmark tasks for other agents. This would create a co-evolutionary dynamic where as agents improve at finding vulnerabilities, the benchmarks they generate become harder, and vice versa.

	Early work in this direction is promising. Automated vulnerability injection into real codebases has been used to create training data for static analysis tools~\cite{DolanGavitt16}, and similar techniques could produce evaluation tasks of calibrated difficulty. The risk is circularity: if the agent generating the benchmark shares biases with the agent being tested, the resulting tasks may become increasingly biased towards the narrow set of vulnerabilities that particular training data induces. Although biases cannot be completely ruled out in a fully generative benchmark, diversity across models, training data, tool configurations, and sampling temperatures is essential to limit this effect.

	\section{Runtime Uncertainty}

	Finally, the actions of AI agents are shaped by multiple sources of runtime uncertainty, each carrying security implications that current benchmarks often fail to capture in their design.

	\begin{runexample}[Self-interference via code]
		To analyze a vulnerability in \ourbench, the agent writes a custom fuzzing harness, which itself contains a buffer overflow. When the harness crashes, the agent attributes the crash to the target program and reports a defect. \ourbench's grading checks only whether the agent triggers a crash in the test environment and marks this as a successful detection. The agent receives credit for finding a vulnerability that exists only in its own code.
	\end{runexample}

	\subsection{Stochastic Behavior}
	Agents built on LLMs are inherently stochastic. At each iteration, an agent may draw on a large variety of actions, tools, and information sources, creating a cascade effect in which early divergence compounds into substantially different outcomes downstream. Consequently, the same agent given the same task can arrive at markedly different conclusions. This non-determinism has direct security implications: an agent that produces a secure solution in one run may produce a vulnerable one in the next. Current benchmarks typically report single-run or mean performance, obscuring the variance that matters most for security evaluation.
	Evaluating stochastic behavior requires running agents multiple times on the same task and reporting distributional statistics: not just mean accuracy but worst-case performance, variance, and the frequency of security-critical failures~\cite{bates2026RLCyber}.

	\subsection{Code Generation}
	Another source of runtime uncertainty is code generation. During a security experiment, an agent may produce a considerable amount of code to probe the environment, create test harnesses, monitor execution, and craft exploits. This code generation underpins agentic security analysis, but it also introduces new points of failure. Since LLM-generated code contains a significant share of security flaws in general, the code an agent produces becomes a crucial variable in the evaluation that most be considered.

	The most direct threat to benchmark validity here is self-interference. An agent may uncover vulnerabilities it introduced itself, or miss flaws it accidentally patched through its own actions, making it impossible to attribute results to the target system rather than the agent's own code. Beyond this, security-relevant properties of the benchmark environment, such as compiler features and memory protection mechanisms, interact with agent-generated code in ways that further confound results. Current benchmarks treat agent-generated code as a transparent instrument, but it is better understood as a source of noise that needs to be controlled for to obtain reliable results.

	\subsection{External Dependencies}

	Effective agents rely on external tools, APIs, and data sources at runtime. A penetration testing agent may query a vulnerability database, invoke a network scanner, or retrieve exploit code from a repository. Each external dependency introduces a potential point of uncertainty: external information may interfere with the analysis process or provide shortcuts to solving the benchmark task, leading to an overestimation of agent capabilities.
	The most critical threat here is information leakage through external sources. Parts of a benchmark solution may be directly accessible via public vulnerability databases, repositories, or online write-ups, either as an explicit leak or as subtle hints that steer the agent toward the correct answer without revealing it outright. In both cases, the agent appears capable when it is largely retrieving rather than reasoning.

	Beyond leakage, external sources may return incorrect or adversarially crafted data, exposing agents to manipulation they are rarely tested against. Current security benchmarks do not control for external information access, making it impossible to completely distinguish agent capability from unintended information gain.

	\subsection{Benchmark Introspection}

	These sources of runtime uncertainty motivate a practical validation technique: \emph{benchmark introspection}. Instead of focusing only on the outcome of a task, the benchmark needs to supervise and validate the evaluation environment through introspection. In the simplest case, this amounts to analyzing the reasoning traces of an agent during the task. In a more advanced form, introspection can externally monitor and document code generation, tool interactions, and filesystem access through appropriate techniques running outside the agent's environment. Similarly, external information can be tainted and tracked through the processing of the agent, providing a clear picture of how information was collected and processed.

	This detailed auditing of the task-solving process, though demanding, can help uncover and mitigate sources of uncertainty. While introspection cannot eliminate the problem entirely, it provides a means for detecting and controlling for the runtime factors that undermine benchmark validity.

	\section{Discussion}

	From a broader perspective, benchmark vulnerabilities, temporal staleness, and runtime uncertainty are not unique to security benchmarks, but they are uniquely acute there. From this, we can draw several conclusions:

	\paragraph{(a) Adversarial reasoning is the capability being measured.} In non-security benchmarks, an agent that games the evaluation is exhibiting undesirable behavior. In security benchmarks, the same capability; finding and exploiting vulnerabilities in its environment, is precisely what we are trying to measure. Cheating and competence are, in this domain, expressions of the same underlying capability. Security benchmarks must be designed accordingly.

	\paragraph{(b) Computer security is non-stationary by nature.} Most benchmarks decay slowly; security benchmarks decay as fast as vulnerabilities are disclosed and patched. A benchmark built on last year's CVEs measures familiarity with last year's problems. Where non-security evaluations may rely on established practices and static data, security benchmarks need to be rethought, not merely refreshed.

	\paragraph{(c) The agent is both the subject and the tool.} The agent generates code, interacts with environments, and makes decisions---all of which have security implications that compound rather than cancel. An output-only evaluation that checks whether the agent found the right vulnerability misses whether it introduced new ones along the way.

	\medskip
	For the practitioner, these points translate into concrete evaluation challenges that no existing benchmark fully addresses:

	\begin{itemize}

		\item \emph{Can the agent attack the benchmark?} Can the agent exploit weaknesses in the infrastructure?

		\item \emph{Is the agent solving today's problems?} Are the benchmark's challenges grounded in vulnerabilities and techniques that remain relevant?

		\item \emph{Is the agent's architecture secure?} Does the agent itself contain exploitable vulnerabilities, in its prompt handling, tool integration, memory management, or privilege boundaries?

		\item \emph{Is the code the agent generates secure?} When agents produce code at runtime is it free of vulnerabilities effecting the benchmark?

	\end{itemize}

	We also note a gap between offensive and defensive evaluation. Current benchmarks focus almost exclusively on offensive capabilities (finding and exploiting vulnerabilities) or on agent safety (resistance to prompt injection and manipulation). No existing benchmark simultaneously evaluates whether an agent is both \emph{effective} at security tasks and \emph{safe} to deploy, yet this is precisely what practitioners need to know.

	\section{Conclusion}

	The rapid development of AI agents for security tasks has outpaced the development of evaluation methodologies adequate to the challenge. We have identified three fundamental classes of vulnerability in current benchmarking approaches and proposed different countermeasures, with benchmark introspection as a holistic first step toward more trustworthy evaluation.

	Much remains open. Generative benchmarks that evolve with the threat landscape, runtime code security evaluation for agent-generated artifacts, and unified offensive-defensive evaluation frameworks all represent important directions for future work. What we hope to have established is that the security community should approach agent benchmarks with the same adversarial skepticism it brings to any other system making security claims---because the benchmarks themselves are not yet secure.

	\section{Acknowledgments}
	This article arose from discussions at the Dagstuhl Perspective Workshop ``Autonomous Agents in Computer Security'' (26162), April 2026. We thank all workshop participants for their contributions. In accordance with IEEE policy, we disclose that Claude Opus~4.6 (Anthropic, 2025) was used to assist with literature search, LaTeX formatting, and copy-editing of drafts. All technical content, analysis, and conclusions are the authors' own.

	\def\refname{References}

\end{multicols}


\begin{thebibliography}{10}
	\small\raggedright

		\bibitem{cybench}
		A.~K.~Zhang et~al., ``Cybench: A framework for evaluating cybersecurity capabilities and risks of language models,'' in \emph{Proc. Int. Conf. Learn. Representations (ICLR)}, 2025.

		\bibitem{cybergym}
		Z.~Wang et~al., ``CyberGym: Evaluating AI agents' real-world cybersecurity capabilities at scale,'' in \emph{Proc. Int. Conf. Learn. Representations (ICLR)}, 2026.

		\bibitem{pentestgpt}
		G.~Deng et~al., ``PentestGPT: Evaluating and harnessing large language models for automated penetration testing,'' in \emph{Proc. 33rd USENIX Security Symp.}, 2024.

		\bibitem{agentauditor}
		H.~Luo et~al., ``AgentAuditor: Human-level safety and security evaluation for LLM agents,'' in \emph{Proc. Advances Neural Inf. Process. Syst. (NeurIPS)}, 2025.

		\bibitem{Tavallaee09}
		M.~Tavallaee et~al., ``A detailed analysis of the KDD CUP 99 data set,'' in Proc. 2nd IEEE Symp. Comput. Intell. Security Defence Appl. (CISDA), 2009.

		\bibitem{fuzzbench21}
		J.~Metzman, L.~Szekeres, L.~M.~R.~Simon, R.~T.~Sprabery, and A.~Arya, ``FuzzBench: An open fuzzer benchmarking platform and service,'' in Proc. 29th ACM Joint Meeting Eur. Softw. Eng. Conf. Symp. Found. Softw. Eng. (ESEC/FSE), 2021, pp. 1393--1403.

		\bibitem{DolanGavitt16}
		B.~Dolan-Gavitt et~al., ``LAVA: Large-scale automated vulnerability addition,'' in Proc. IEEE Symp. Security Privacy (S\&P), 2016, pp. 110--121.

		\bibitem{berkeley}
		Y.~Zhu et~al., ``Establishing best practices for building rigorous agentic benchmarks,'' arXiv:2507.02825, 2025.

		\bibitem{secureagentbench}
		J.~Chen et~al., ``SecureAgentBench: Benchmarking secure code generation under realistic vulnerability scenarios,'' arXiv:2509.22097, 2025.

		\bibitem{berkeley}
		Hao Wang et~al., ``How We Broke Top AI Agent Benchmarks: And What Comes Next'', 2026. \href{https://rdi.berkeley.edu/blog/trustworthy-benchmarks-cont/}{{\footnotesize\textit{https://rdi.berkeley.edu/blog/trustworthy-benchmarks-cont/}}}

		\bibitem{browsecomp}
		Anthropic, ``Eval awareness in Claude Opus 4.6's BrowseComp performance'', 2026. \href{https://www.anthropic.com/engineering/eval-awareness-browsecomp}{{\footnotesize\textit{https://www.anthropic.com/engineering/eval-awareness-browsecomp}}}

		\bibitem{bates2026RLCyber}
		Bates et al., ``Beyond Rewards in Reinforcement Learning for Cyber Defence'', ICML, 2026

		\bibitem{satmlmi}
		Zhang et al., ``Membership Inference Attacks Cannot Prove that a Model Was Trained On Your Data,'' in Proc. IEEE SaTML, 2025.

		\bibitem{hardtbench}
		Moritz Hardt, ``The Emerging Science of Machine Learning Benchmarks'', Princeton University Press, 2026.

	\end{thebibliography}
\end{document}